\documentstyle[twocolumn,aps,psfig]{revtex}

\begin{document}
\draft
\flushbottom
\twocolumn[
\hsize\textwidth\columnwidth\hsize\csname @twocolumnfalse\endcsname

\title{ Vacuum diamagnetism of mesoscopic metallic samples
}
\author{I. I. Smolyaninov}
\address{ Electrical and Computer Engineering Department,
University of Maryland, College Park, MD 20742}
\date{\today}
\maketitle
\tightenlines
\widetext
\advance\leftskip by 57pt
\advance\rightskip by 57pt
\begin{abstract}

Zero-point energy of surface plasmon modes of mesoscopic metal samples in an external magnetic field has been considered. The magnetic response of plasmon vacuum has shown to be diamagnetic. In thin films this surface vacuum diamagnetism is at least of the same order of magnitude as the magnetism of the bulk electrons. Thus, a novel type of magnetism shown by thin metallic samples, which is complementary to the well known Pauli paramagnetic and Landau diamagnetic contributions to the magnetism of the electron gas has been demonstrated. 


\end{abstract}

]

\narrowtext
\tightenlines

Transport and magnetic properties of mesoscopic systems are the topics of very active current research, due to the strong drive towards development of novel nanodevices and quantum computing. However, there still exist
numerous current contradictions between mesoscopic theories and experiment \cite{1,2,3}. While a lot of attention is concentrated on the potential role of sample imperfections such as magnetic impurities, at least some of these contradictions may be explained by previously unaccounted intrinsic mechanisms. 
One of the missing intrinsic elements in many current experimental and theoretical studies is the effect of surface plasmons \cite{4} on the transport and magnetic properties of mesoscopic metallic samples, such as wires, rings, and various other shapes. Importance of surface plasmons in mesoscopic phenomena is clear from the recent paper \cite{5} and from the following qualitative consideration. 

The usual justification for not considering surface plasmons in low temperature measurements is that surface plasmons are not excited if the sample size is of the order of a few micrometers and the temperature is low. However, even if there are no real plasmon quanta in the system, the zero-point fluctuations of the electromagnetic field of all the possible plasmon modes in the system have to be considered. The importance of zero-point fluctuations in mesoscopic systems clearly manifests itself in the observations of negative vacuum energy density between metal plates separated by submicrometer distances (the Casimir effect, see for example \cite{6}). The energy density in such a mesoscopic cavity depends on the dielectric constant of the material between the metal plates. A mesoscopic metallic sample constitutes a similar mesoscopic resonator for surface plasmons. As has been shown in numerous papers, magnetic field produces substantial modifications of the surface plasmon dispersion law via modification of the dielectric tensor of the metal \cite{7}, and via the Aharonov-Bohm (AB) effect (the frequencies of surface plasmon modes in nanotubes and mesoscopic rings change periodically by about 10 percent due to the AB effect \cite{8}). Thus, magnetic field applied to a mesoscopic sample effectively changes its "dielectric constant" as seen by the surface plasmons, and hence, changes the zero-point energy of the plasmon field. This fact may be interpreted as an additional vacuum contribution to the magnetic moment $\mu _{vac}=-\partial E_0/\partial H$ of the metal sample. According to the results presented in \cite{7,8}, this vacuum contribution may be quite large, because of the rather large magnetic moments $-\hbar \partial \omega/\partial H \sim \mu _B$ of the individual plasmon modes. 

In this Letter I am going to calculate the plasmon zero-point energy of some metallic mesoscopic samples with simple geometries in an external magnetic field. The magnetic response of these plasmon vacuums appears to be rather large and diamagnetic. In thin films this predicted novel surface vacuum diamagnetism appears to be at least of the same order of magnitude as the magnetism of the bulk electrons. Thus, a novel type of magnetism shown by thin metallic samples, which is complementary to the well known Pauli paramagnetic and Landau diamagnetic contributions to the magnetism of electron gas will be demonstrated. 

Let us consider the magnetic response of the surface plasmon vacuum of a square $a\times a$ region of a thin metal film with thickness $d<<a$ to an applied perpendicular magnetic field $H$ (Fig.1). Let us start by considering the dispersion law of a surface plasmon (SP), which propagates along the metal-dielectric interface in zero magnetic field. The SP field decays exponentially both inside the metal and the dielectric. Let us assume that both metal and dielectric completely fill the respective $z<0$ and $z>0$ half-spaces. In such a case the dispersion law can be written as \cite{4} 

\begin{equation}  
k^2=\frac{\omega ^2}{c^2}\frac{\epsilon _d\epsilon _m(\omega )}{\epsilon _d+\epsilon _m(\omega)} ,
\end{equation}

where we will assume that $\epsilon _m=1-\omega _p^2/\omega ^2$ according to the Drude model, and $\omega _p$ is the plasma frequency of the metal. This dispersion law is shown in Fig.2 for the cases of metal-vacuum and metal-dielectric interfaces. It starts as a "light line" in the respective dielectric at low frequencies and approaches asymptotically $\omega =\omega _p/(1+\epsilon _d)^{1/2}$ at very large wave vectors. The latter frequency corresponds to the so-called surface plasmon resonance. Under the surface plasmon resonance conditions both phase and group velocity of the SPs is zero, and the surface charge and the normal component of the electric field diverge. Since at every wavevector the SP dispersion law is located to the right of the "light line", the SPs of the plane metal-dielectric interface are decoupled from the free-space photons due to the momentum conservation law.   

We are mostly interested in the region of the dispersion law near the plasmon resonance since our samples of interest are "mesoscopic", and hence $a$ is much smaller than $2\pi c/\omega _{sp}$, where $\omega _{sp}$ is the surface plasmon resonance frequency. The corrections to surface plasmon dispersion law in an applied magnetic field were calculated by Chiu and Quinn \cite{7}. Let us initially consider the most simple case of magnetic field $H$ applied perpendicular to the metal film bounded by vacuum. At large plasmon wavevectors $k$ the surface plasmon frequency 

\begin{equation}
\omega _{sp} \approx ((\omega _p^2+\omega _c^2)/2)^{1/2} 
\end{equation}

does not depend on $k$ (Fig.2), where $\omega _p$ and $\omega _c=eH/mc$ are the plasma and cyclotron frequencies, respectively \cite{7}. As a result, all the surface plasmon modes have the same magnetic moment $-\hbar \partial \omega/\partial H$, and the total magnetic moment of the plasmon vacuum can be written as 

\begin{equation}
\mu _{vac} \approx - \mu _B/2^{1/2} \Sigma _k \omega _c/\omega _p , 
\end{equation}

where $\mu _B$ is the Bohr magneton, and summation has to be done over all the plasmon modes of the square region of the thin metal film under consideration.
The surface plasmon eigenmodes of this square region are defined by the two-component wave vector $(k_x,k_y)=\pi/a\times (n_x,n_y)$, where $n_x$ and $n_y$ are integer. Due to Landau damping \cite{9} the summation over all possible surface plasmon wave vectors has to be cut off at $\mid k_{max}\mid \sim k_F$ (the electron Fermi momentum). Thus, the total number of plasmon modes on the top and bottom interfaces of the metal film is roughly $8(k_Fa/\pi )^2$, and 

\begin{equation}
\mu _{vac}\approx -8a^2n_e^{2/3}(\omega _c/\omega _p)\mu _B 
\end{equation}

For thin metal films this surface vacuum diamagnetism is at least of the same order of magnitude as the contribution of the bulk electrons. Detailed description of various contributions to the magnetism of metallic samples can be found, for example in \cite{10}. The dominating mechanism of the magnetism of conductivity electrons is the paramagnetic contribution first obtained by Pauli \cite{11} as

\begin{equation}
\chi _e^{pm}=\frac{3^{1/3}m\mu _B^2}{\pi ^{4/3}\hbar ^2}n_e^{1/3} ,
\end{equation} 

where $m$ is the electron mass. The diamagnetic contribution first obtained by Landau \cite{12} is usually smaller by approximately a factor of 3 \cite{10}. Thus, we only need to compare the magnitudes of the Pauli paramagnetic contribution and the surface plasmon vacuum contribution to the electron magnetism of our thin metal film sample. Let us assume the free-electron model value for the plasma frequency of electron gas in the metal \cite{4}:

\begin{equation}
\omega _p^2=\frac{4\pi e^2n_e}{m} ,
\end{equation}

and compare these two contributions. The Pauli paramagnetic moment of our sample can be written as follows:

\begin{equation}
\mu _{pm}\approx a^2n_e^{2/3}\omega _c\frac{3^{1/3}md}{2\pi ^{4/3}\hbar n_e^{1/3}}\mu _B = 8a^2n_e^{2/3}(\omega _c/\omega _d)\mu _B,
\end{equation}

where $d$ is the film thickness, and some characteristic frequency 
$\omega _d=16\pi ^{4/3}\hbar n_e^{1/3}/(3^{1/3}md)$ is introduced. From this expression we immediately see that at small film thicknesses $d$, such that $\omega _p\sim \omega _d$, Pauli paramagnetic and surface plasmon vacuum diamagnetic contributions have the same order of magnitude. The characteristic film thickness necessary for this situation to occur can be written as

\begin{equation}
d\sim \frac{16\pi ^{4/3}\hbar n_e^{1/3}}{3^{1/3}m\omega _p}\sim \frac{8\pi ^{5/6}\hbar }{3^{1/3}m^{1/2}en_e^{1/6}} \sim 2nm
\end{equation}

Thus, according to this simple estimate Pauli paramagnetic and surface plasmon vacuum diamagnetic contributions to the magnetism of a thin film sample would be approximately equal at $d\sim 2$nm. Moreover, the contributions of Landau diamagnetism and plasmon vacuum diamagnetism should be about the same at $d\sim 6$nm. As a result, experimental measurements of the magnetic response of mesoscopic thin film samples performed as a function of film thickness in the 2-20 nm range have reasonable chance of detecting the surface plasmon vacuum diamagnetism. I should also point out that in reality the vacuum diamagnetic moment may be at least an order of magnitude larger than the value determined by equation (4), since the cut-off wave vector is known only by an order of magnitude. In addition, magnetic response measurements may be performed as a function of the dielectric constant of the substrate and/or absorbed layer on the interfaces of the metal film. According to Chiu and Quinn \cite{7}, in the presence of a dielectric layer on the metal surface and at large plasmon wavevectors $k$ the surface plasmon frequency in perpendicular magnetic field is given by

\begin{equation}
\omega _{sp} \approx ((\epsilon _d^{-1}\omega _p^2+\omega _c^2)/2)^{1/2} ,
\end{equation}
  
where $\epsilon _d$ is the dielectric constant of the layer. As a result, the magnetic moment of individual plasmon modes $-\hbar \partial \omega/\partial H$, and the total magnetic moment of the plasmon vacuum (defined by equations (3) and (4)) are multiplied by the refractive index of the dielectric $n_d=\epsilon _d^{1/2}$. Since atomic monolayer quantities of the adsorbents are sufficient to shift the plasmon resonance \cite{4}, such measurements may be very useful in separating the relative contribution of the vacuum diamagnetism into the overall magnetic response of the sample (since there is no reason why a thin absorbed layer would alter either Pauli or Landau contributions). In addition, increase of the overall effect by another factor of 2 or 3 means that even thicker metal films (few tens of nanometers) may be used to make a mesoscopic sample of interest. This additional increase in the scale of the film thickness also means that calculations of plasmon eigenfrequencies based on the macroscopic Maxwell equations are much more reliable.   

Except for the extreme sensitivity to the absorbed layers, the vacuum diamagnetism described above looks similar to the Pauli and Landau contributions with respect to temperature changes. Since surface plasmon eigenfrequencies involved are of the order of a few electron-Volts, no considerable changes in the surface plasmon magnetic response may be expected from absolute zero up to the room temperature. The best candidates for vacuum diamagnetism observations may be the mesoscopic samples made of gold, silver, copper or aluminum, since these metals exhibit very pronounced plasmon resonances \cite{4}. Even in the presence of absorbed layers the vacuum diamagnetic contribution defined by equation (4) is rather small due to the small $(\omega _c/\omega _p)$ factor.
For a gold sample at $H=1$T this ratio is $(\omega _c/\omega _p)\sim 10^{-4}$. Thus, a 1 micrometer by 1 micrometer square sample would have vacuum diamagnetic moment of $\sim 10^4\mu _B$ at $H=1$T , or a few tens of Bohr magnetons at $H=10$G. However, similar sensitivity has been achieved recently by Deblock $et$ $al.$ \cite{2} who reported measurements of the magnetic response of individual mesoscopic silver rings and obtained that the measured response of an individual silver ring is diamagnetic in the limit of zero magnetic field, which was not consistent with the available theoretical predictions. According to these measurements performed on an ensemble of $1.5\times 10^5$ 1 micrometer by 1 micrometer square silver rings with the thickness of 70 nm, the magnetic moment of an individual ring oscillates with the number of magnetic flux quanta with an amplitude of approximately $30\mu _B$ and the period of $\sim 20$G. Regardless of the nature of these oscillations, which we will briefly discuss below, the result of Deblock $et$ $al.$ for the unexpected diamagnetic response of an individual ring seams consistent with both the sign and the magnitude (according to Fig.4 from \cite{2}, $\mu _{ring}\sim 10\mu _B$) of the vacuum diamagnetic contribution described above. 

Nevertheless, the results of \cite{2} should be taken with some degree of caution, because of the chosen experimental geometry. Looking at the geometry of the experiment presented in the Fig.1 of \cite{2}, we immediately observe that the plasmon modes of individual rings in the sample are strongly coupled to each other. The surface plasmon wavelengths of the individual rings may be roughly estimated as $\lambda _p\sim L/n$ where $L=4 \mu m$ is the ring perimeter, and $n$ is an integer. This corresponds to the plasmon field decay outside a ring with an exponent $\lambda _p/2\pi \sim 640/n$ nm. On the other hand, the distance between the individual rings in the periodic array shown in Fig.1 of \cite{2} appears to be $\sim 550 nm$. As a result, the plasmon modes of the individual rings are coupled to each other, and the calculation of surface plasmon spectrum of an array of such coupled cylindrical rings represents rather difficult problem. The fact that the rings are coupled indicates that Deblock $et$ $al.$ may not have achieved proper measurement of the magnetic response of an individual mesoscopic silver ring. The proper measurement can be done if the spacing of the individual rings will be made substantially larger than the ring perimeter, so that individual plasmon modes will be decoupled.  

In some geometries the frequencies of individual plasmon modes may exhibit linear dependence on the applied magnetic field \cite{7,8,13,14}. For example, if magnetic field is applied parallel to the metal film (in the $x$-direction), than according to Chiu and Quinn \cite{7} at large plasmon wavevectors $k$ the surface plasmon frequency is given by

\begin{equation}
\omega _{sp} \approx ((\omega _p^2+\frac{1}{2}\omega _c^2)/2)^{1/2}\pm \frac{1}{2}\omega _c , 
\end{equation}

where the two signs correspond to plasmon propagation in the $\mp y$-direction. Such plasmon modes have magnetic moments $-\hbar \partial \omega/\partial H\sim \pm \mu _B$. In a similar way, cylindrical surface plasmon modes of nanoholes and nanotubes \cite{8,14} with nonzero angular momenta also have magnetic moments $\sim \pm \mu _B$ \cite{14}. In rotationally symmetric samples vacuum magnetic moments of these \lq\lq left \lq\lq and \lq\lq right \lq\lq modes compensate each other, so that only the contributions quadratic in the magnetic field are left in the expression for the total zero-point energy of the plasmon vacuum, and we would come up with an expression more or less similar to equation (4) for the vacuum magnetic moment. However, this may not be the case for an asymmetric sample. While surface plasmon modes with very large wave vectors may not be affected much by the asymmetry of the sample shape, small number of low $k$ modes will be very sensitive to it. A small number of such \lq\lq asymmetric \lq\lq plasmon modes may be responsible for the diamagnetic moment $\sim 10\mu _B$ observed by Deblock $et$ $al.$ in zero magnetic field if the rings used in their experiment are not perfectly symmetric. In addition, asymmetry between the \lq\lq left \lq\lq and \lq\lq right \lq\lq plasmon modes may be caused by natural or magnetic field induced optical activity \cite{15} of the mesoscopic sample itself and/or the absorbed layer on its surface. Optical activity of materials and samples is usually described by the difference in the refractive indices $n_+$ and $n_-$ for the left and right circular polarizations (in other words, modes with opposite angular momenta) of light. Because of the different refractive index, an effective optical length of an optically active cylindrical sample is different for the left and right plasmon modes. As a result, the total number of the left and right plasmon modes may be slightly different, and the cylinder may posses an additional vacuum magnetic moment

\begin{equation}
\mu _{vac}\sim (n_+ - n_-)han_e^{2/3}\mu _B ,
\end{equation}
  
where $h$ is the height of the cylinder and $a$ is its radius. Here we should also point out that all metals exhibit magnetic field induced optical activity \cite{15}, so that this effect may be also quite relevant for the mesoscopic metallic samples in an external magnetic field. 

Another potentially important (although separate) question is how the Aharonov-Bohm effect may affect the vacuum diamagnetism of mesoscopic samples. Very recently Chaplik $et$ $al.$ demonstrated that the frequencies of surface plasmon modes in nanotubes and mesoscopic rings change periodically by about 10 percent due to the AB effect \cite{8}. The reason for this effect is the fact that both Fermi energy and the polarization operator of the nanotubes and nanorings changes periodically with magnetic flux due to the periodic dependence of the single-particle spectrum on magnetic flux. These periodic changes lead to the periodic changes of the surface plasmon eigenfrequencies, which can be found by solving the Poisson equation. Thus, magnetic moments of these modes $-\hbar \partial \omega/\partial H$ experience periodic oscillations too. According to the results of numerical calculations by Chaplik $et$ $al.$ \cite{8} the amplitude of these oscillations is of the order of $\mu _B$ for the zero-angular momentum ($m=0$) plasmon mode of a carbon nanotube at $ka=1$ (Fig.2(b) from \cite{8}), while this mode does not have any angular momentum in zero magnetic field (for this mode $d\omega /dH=0$ at $H=0$). Similar to the classical result for cylindrical surface plasmons with nonzero angular momentum, plasmons of the nanotube with $m\neq 0$ have nonzero AB-induced magnetic moments (Fig.4 from \cite{8}). While exact numerical calculations for the total AB-induced vacuum magnetic moment would be rather cumbersome, because of the complicated dependencies of the frequencies of the plasmon modes with arbitrary $m$ and $k$ on the magnetic flux \cite{8}, there is absolutely no reason for the total magnetic moment not to experience periodic oscillations with the amplitude of a few $\mu _B$ due to the AB effect. Thus, the periodic small oscillations of the magnetic response of the silver rings observed by Deblock $et$ $al.$ \cite{2} may also be at least partially attributed to the oscillations of vacuum magnetism. 

In conclusion, zero-point energy of surface plasmon modes of mesoscopic metal samples in an external magnetic field has been considered. The magnetic response of plasmon vacuum has shown to be diamagnetic. In thin films this surface vacuum diamagnetism is at least of the same order of magnitude as the magnetism of the bulk electrons. Thus, a novel type of magnetism shown by thin metallic samples, which is complementary to the well known Pauli paramagnetic and Landau diamagnetic contributions to the magnetism of the electron gas has been demonstrated. 

This work has been supported in part by the NSF grants ECS-0210438 and ECS-0304046.

Figure captions.

Fig.1 Model geometry for the calculations of the magnetic response of surface plasmon vacuum in the case of a square $a\times a$ region of a thin metal film with thickness $d<<a$ in an applied perpendicular magnetic field $H$.

Fig.2 Surface plasmon dispersion law for the cases of metal-vacuum and metal-dielectric interfaces in zero magnetic field (solid lines) and in the magnetic field applied perpendicular to the film surface (dashed lines). At large plasmon wavevectors $k$ the changes in surface plasmon frequency in the applied magnetic field do not depend on $k$.

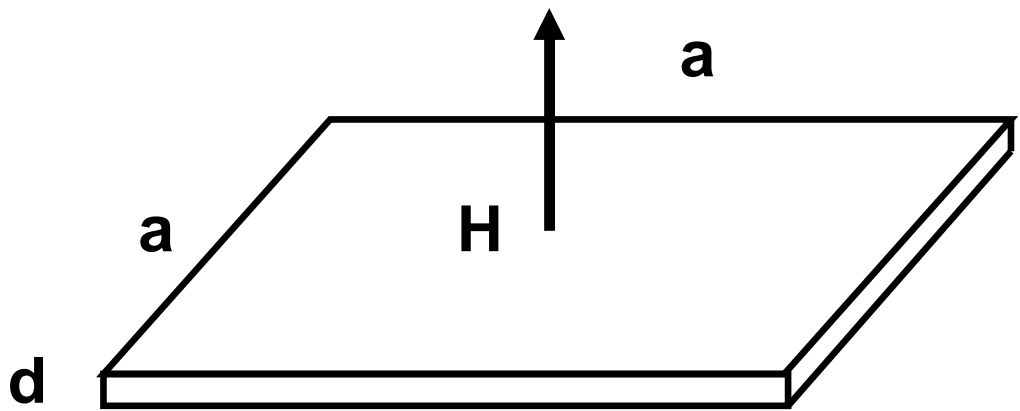
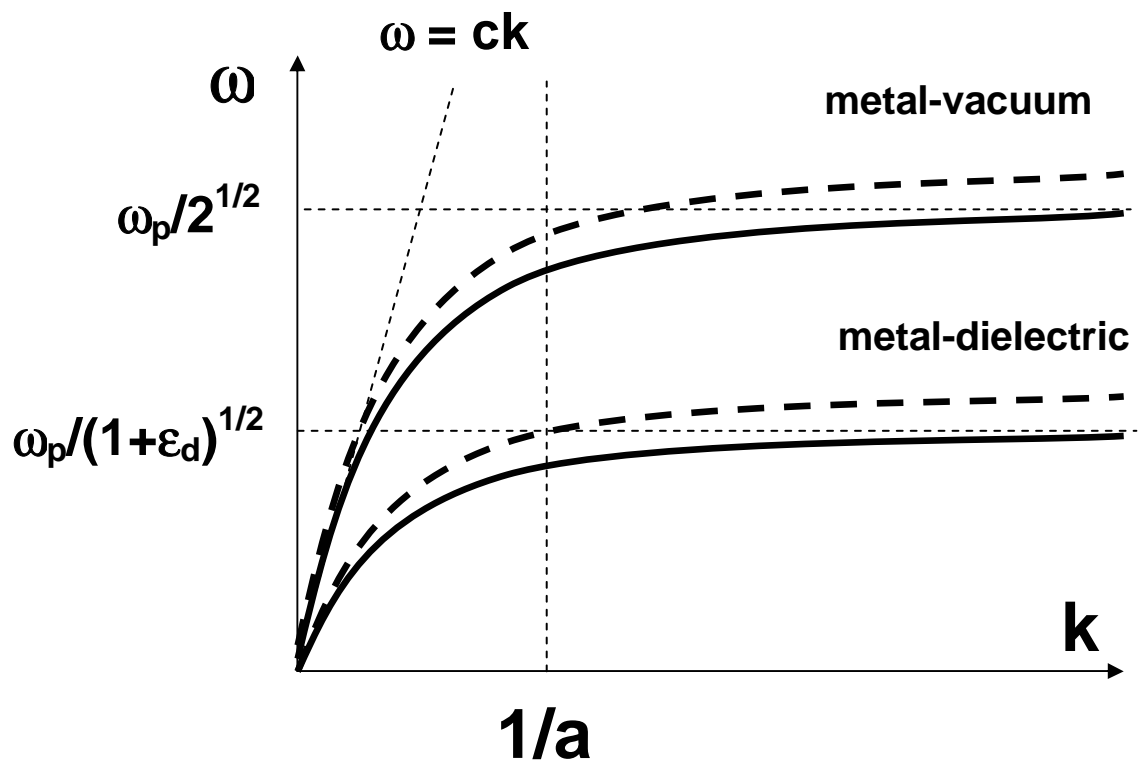
\end{document}